\documentclass[11pt]{article}
\usepackage{amssymb,amsmath}
\usepackage{float}
\usepackage{color,fullpage}
\usepackage{amsthm}
\usepackage{graphicx}   
\usepackage{algorithm}
\usepackage{algorithmic} 
\usepackage{graphics} 
\usepackage{caption}
\usepackage{subcaption}

\newtheorem{theorem}{Theorem}

\newtheorem{lemma}{Lemma}

\newtheorem{definition}{Definition}

\numberwithin{equation}{section}
\numberwithin{figure}{section}
\numberwithin{definition}{section}

\begin{document}
\bibliographystyle{unsrt}
\title{Phase Retrieval Via Reweighted Wirtinger Flow}

\author{Ziyang Yuan\textsuperscript{a}\thanks{
College of Science, National University of Defense Technology,
Changsha, Hunan, 410073, P.R.China. Corresponding author. Email: \texttt{yuanziyang11@nudt.edu.cn}}
\and Hongxia Wang{\textsuperscript{a}\thanks{
		College of Science, National University of Defense Technology,
		Changsha, Hunan, 410073, P.R.China. Corresponding author. Email: \texttt{whx8292@hotmail.com}}
}
}



\date{}

\maketitle

\begin{abstract}
	Phase retrieval(PR) problem is a kind of ill-condition inverse problem which is arising in various of applications. Based on the Wirtinger flow(WF) method, a reweighted Wirtinger flow(RWF) method is proposed to deal with PR problem. RWF finds the global optimum by solving a series of sub-PR problems with changing weights. Theoretical analyses illustrate that the RWF has a geometric convergence from a deliberate initialization when the weights are bounded by 1 and $\frac{10}{9}$. Numerical testing shows RWF has a lower sampling complexity compared with WF. As an essentially adaptive truncated Wirtinger flow(TWF) method, RWF performs better than TWF especially when the ratio between sampling number $m$ and length of signal $n$ is small.\\
	$\mathbf{keywords}$:~~Phase retrieval~~Wirtinger flow~~Gradient descent~~Reweighted
	
\end{abstract}

\section{Introduction}
\label{intro}
\subsection{Phase retrieval problem}
\indent 
In optics, most detectors can only record the intensity of the signal while losing the information about the phase. Recovering the signal from the amplitude only measurements is called phase retrieval problem(PR) which arises in a wild range of applications such as Fourier Ptychography Microscopy, diffraction imaging, X-ray crystallography and so on\cite{Bunk2007Diffractive}\cite{Miao1999Extending}\cite{Zheng2013Wide}. Phase retrieval problem can be an instance of solving a system of quadratic equations: 
\begin{eqnarray}
\centering
y_i=|\langle\mathbf{a}_i,\mathbf{x}\rangle|^2+\mathbf{\varepsilon}_i,~~i=1,...,m,
\end{eqnarray}
where $\mathbf{x}\in\mathbb{C}^n$ is the signal of interest, $\mathbf{a}_i\in\mathbb{C}^n$ is the measurement vector, $y_i\in \mathbb{R}$ is the observed measurement, $\varepsilon_i$ is the noise.\\ 
\indent
(1.1) is a non-convex and NP-hard problem. Traditional methods usually fall to find the solutions. Besides, let $\tilde{\mathbf{x}}$ be the solution of (1.1), Obviously, $\tilde{\mathbf{x}}e^{i\theta}$ also satisfies (1.1) for any $\theta\in[0,2\pi]$. So the uniqueness of the solution to (1.1) is often defined up to a global phase factor.
\subsection{Prior art}
For classical PR problem, $\{\mathbf{a}_i\}_{1\leq i\leq m}$ are the Fourier measurement vectors. There were series of methods came up to solve (1.1). In 1970, error reduction methods such as Gerchberg-Saxton and Hybrid input and output method\cite{Fienup1982Phase}\cite{Gerchberg1971A} were proposed to deal with phase retrieval problems by constantly projecting the evaluations between transform domain and spatial domain with some special constraints. These methods often get stuck into the local minimums, besides, fundamental mathematical questions concerning their convergence remain unsolved. In fact, without any additional assumption over $\mathbf{x}$, it is hard to recover $\mathbf{x}$ from $\{\mathbf{y}_i\}_{1\leq i\leq m}$. For Fourier measurement vectors, the trivial ambiguities of (1.1) are including global phase shift, conjugate inversion and spatial shift. In fact, it has been proven that 1D Fourier phase retrieval problem has no unique solution even excluding those trivialities above. To relief the ill condition characters, one way is to utilize some priori conditions such as nonnegativity and sparsity. Gespar\cite{Shechtman2013GESPAR} and dictionary learning method\cite{tillmann2016dolphin} both made progress in these regions.\\
\indent
With the development of compress sensing and theories of random matrix, measurement vectors $\{\mathbf{a}_i\}_{1\leq i\leq m}$ aren't merely constrained in a determined type. When $m\geq c_0n\mathrm{log}n$ and $\mathbf{a}_i\overset{i.i.d}\sim\mathcal{N}(\mathbf{0},\mathbf{I})$, Wirtinger flow(WF)\cite{candes2015phase} method can efficiently find the global optimum of (1.1) with a careful initialization. The objective function of the WF is a forth degree smooth model which can be described as below:
\begin{eqnarray}
\mathop {\mathrm{minimize}}\limits_{\mathbf{z}\in\mathbb{C}^n}~~\mathit{f}(\mathbf{z})=\frac{1}{2m}\sum_{i=1}^{m}(|\langle\mathbf{a}_i,\mathbf{z}\rangle|^2-y_i)^2,
\end{eqnarray}
where $y_i=|\langle\mathbf{a}_i,\mathbf{x}\rangle|^2$. $\mathbf{x}$ is the signal to be reconstructed.\\
\indent
WF evaluates a good initialization by power method and utilizes the gradient descent algorithm in each step with Wirtinger derivative to solve (1.2). It has been established using the theorem of algebra that for real signal, $2n-1$ random measurements guarantee uniqueness with a high probability\cite{Balan2006On}, for complexity signal $4n-4$ generic measurements are sufficient\cite{Bandeira2013Saving}. In general, when $m/n\geq 4.5$, WF can yield a stable empirical successful rate($\geqslant$ 95\%). However when $m/n\leq3$, WF has a low recovery rate. There is a gap between the sampling complexity of the WF and the known information-limit. Along this line, a batch of similar works have been sprung up. In \cite{chen2015solving}, Chen et al. suggest a  truncated Wirtinger flow(TWF) method based on Possion model. TWF can largely improve performance of the WF by truncating some weird components. Zhang et al. came up with a reshaped Wirtinger flow model which have the type below\cite{zhang2016reshaped}:
\begin{eqnarray}
\mathop {\mathrm{minimize}}\limits_{\mathbf{z}\in\mathbb{C}^n}~~\mathit{f}(\mathbf{z})=\frac{1}{2m}\sum_{i=1}^{m}(|\langle\mathbf{a}_i,\mathbf{z}\rangle|-\sqrt{y_i})^2.
\end{eqnarray}
(1.3) is a low-order model compared to (1.2). Though it is not differentiable in those points in $\big\{\mathbf{z}|\mathbf{a}^*_i\mathbf{z}=0,i\in\{1,...,m\}\big\}$, it has little influence on the convergence analysis near the optimal points. Reshaped WF utilizes a kind of sub-gradient algorithm to search for the global minimums. Numerical tests demonstrated its comparative lower sample complexity. For real signal, it can have a 100\% successful recovery rate when $m/n\approx3.8$, for complex signal, the value is about $m/n=4.2$. In \cite{wang2016solving}, it came up with a truncated reshaped WF which decreased the sampling complexity further. Recently, a stochastic gradient descent algorithm was also proposed based on this reshaped model for the large scale problem\cite{wang2016solving1}.
\subsection{Algorithm in this paper}
\indent
In this paper, a reweighted Wirtinger flow(RWF) algorithm is proposed to deal with the phase retrieval problem. It is based on the high-degree model (1.2), but it can have a lower sampling complexity. The weights of every compositions are changing during the iterations in RWF. These weights can have a truncation effect indirectly when the current evaluation is far away from the optimum. Theoretical analysis also shows that once the weights are bounded by 1 and $\frac{10}{9}$, RWF will converge to the global optimum exponentially.\\
\indent
The remainders of this paper are organized as follows. In section 2, we introduce proposed RWF and establish its theoretical performance. In section 3, numerical tests compare RWF with state-of-the-art approaches. Section 4 is the conclusion. Technical details can be found in Appendix.\\
\indent 
In this article, the bold capital uppercase and lowercase letters represent matrices and vectors. $(\cdot)^*$ denotes the conjugate transpose, $j=\sqrt{-1}$ is the imaginary unit. $\mathrm{Re}(\cdot)$ is the real part of a complex number. $|\cdot|$ denotes the absolute value of a real number or the module of a complex number. $||\cdot||$ is the Euclidean norm of a vector.
\section{Reweighted Wirtinger Flow}
\subsection{Algorithm of RWF}
Reweighting skills have sprung up in several relating arts. In \cite{chartrand2008iteratively}\cite{daubechies2010iteratively}, iteratively reweighted least square algorithms were came up to deal with problem in compress sensing. Utilizing the same idea, we suggest a reweighted wirtinger flow model as:
\begin{eqnarray}
\mathop {\mathrm{minimize}}\limits_{\mathbf{z}\in\mathbb{C}^n}~~\mathit{f}(\mathbf{z})=\frac{1}{2m}\sum_{i=1}^{m}f_i(\mathbf{z})=\frac{1}{2m}\sum_{i=1}^{m}\omega_i(|\langle\mathbf{a}_i,\mathbf{z}\rangle|^2-y_i)^2,
\end{eqnarray}
where $\omega_i\geqslant 0$ are weights. We can easily conclude that $\mathbf{x}$ is the global minimum of (2.1). WF is actually a special case of RWF where $\omega_i=1$ for $i=1,...,m$. If $\{\omega_i\}_{1\leq i\leq m}$ are determined, (2.1) can be solved by gradient descent method with Wirtinger gradient $\nabla\mathit{f}(\mathbf{z})$:
\begin{eqnarray}
\nabla \mathit{f}(\mathbf{z})=\frac{1}{m}\sum_{i=1}^{m}\nabla f_i(\mathbf{z})=\frac{1}{m}\sum_{i=1}^{m}\omega_i(|\langle\mathbf{a}_i,\mathbf{z}\rangle|^2-y_i)\mathbf{a}_i\mathbf{a}_i^*\mathbf{z}.
\end{eqnarray}
The details of Wirtinger derivatives can be found in \cite{candes2015phase}.\\
\indent
The key of our algorithm is to determine $\omega_i$. In TWF \cite{chen2015solving}, it sets a parameter $C$ to truncate those $(|\langle\mathbf{a}_i,\mathbf{z}\rangle|^2-y_i)\mathbf{a}_i\mathbf{a}_i^*\mathbf{z}$ where $i\in W(i)=\{i~\Big|\big||\langle\mathbf{a}_i,\mathbf{x}\rangle|^2-y_i\big|\geq C,i=1,...m\}$. Those components may make  $\nabla\mathit{f}(\mathbf{z})$ deviate from the correct direction. To alleviate the selection of $C$, we choose some special $\omega_i$ which can be seen in (2.3). Those weights are adaptively calculated from the algorithm depending on the value of $\mathbf{z}_{k-1}$. From (2.3), we can see that the corresponding $\omega_i$ will be comparatively low when $i\in W(i)$. This small $\omega_i$ will diminish the contribution of $\nabla f_i(\mathbf{z})$ to the $\nabla f(\mathbf{z})$. Thus adding weights is actually an indirect adaptive truncation.
\begin{eqnarray}
\omega_i^k=\frac{1}{\Big|\big|\langle\mathbf{a}_i,\mathbf{z}_{k-1}\rangle\big|^2-y_i\Big|+\eta_i},~i=1,...,m,
\end{eqnarray}
where $\mathbf{z}_{k-1}$ is the result in the (k-1)th iteration. $\eta_i$ is a parameter which can change during the iteration or to be stagnated all the time.\\
\indent 
Then, we design an algorithm called RWF to search for the global minimum $\mathbf{x}$. RWF updates $\mathbf{z}_k$ from a proper initialization $\mathbf{z}_0$ which is caculated by the power method in \cite{candes2015phase}. The details of RWF can be seen in algorithm 1. \\
\begin{algorithm}
	\caption{\textbf{Reweighted Wirtinger Flow}($\mathbf{RWF}$)} 
	\label{alg:Framwork} %
	\renewcommand{\algorithmicrequire}{\textbf{Input:}}
	\renewcommand\algorithmicensure {\textbf{Output:} }
	\begin{algorithmic}   
		\REQUIRE$\{\{y_i\}_{1\leq i\leq m},\{\mathbf{a}_i\}_{1\leq i\leq m},\{\eta_i\}_{1\leq i\leq m}\ ,T\}$ 
		\STATE 
		$\{\mathbf{a}_i\}_{i=1}^m$: Gaussian vectors\\
		$y_i=|\langle\mathbf{a}_i,\mathbf{x}\rangle|^2$: measurements\\
		$\eta_i$: the parameter\\
		$T$: the maximum iteration times\\
		\ENSURE$\mathbf{x}^*$
		\STATE $\mathbf{x}^*$: the reconstructed signal\\
		\vskip 4mm
		\hrule
		\vskip 2mm
		$\mathbf{Initialization}$
		\STATE
		set $\lambda^2=n\frac{\sum_ry_i}{\sum_i||\mathbf{a}_i||^2}$\\
		set $\mathbf{z}_0$, $||\mathbf{z}_0||=\lambda$ to be the eigenvector corresponding to the largest eigenvalue of
		\begin{eqnarray*}
			\mathbf{Y}=\frac{1}{m}\sum_{i=1}^{m}y_i\mathbf{a}_i\mathbf{a}_i^*
		\end{eqnarray*} 
		\vskip 2mm
		\FOR{$k=1$; $k\leq T$; $k++$ }
		\STATE
		$\mathbf{z}_{k+1}\in \mathrm{argmin}f^k(\mathbf{z})$ 
		\ENDFOR\\
		$\mathbf{x}^*=\mathbf{z}^T$
	\end{algorithmic}  
\end{algorithm}
\begin{algorithm}
	\caption{\textbf{Gradient descent method solver of (2.4)}} 
	\label{alg:Framwork} %
	\renewcommand{\algorithmicrequire}{\textbf{Input:}}
	\renewcommand\algorithmicensure {\textbf{Output:} }
	\begin{algorithmic}   
		\REQUIRE$\{\mathbf{z}_k,\{y_i\}_{1\leq i\leq m},\{\mathbf{a}_i\}_{1\leq i\leq m},\{\eta_i\}_{1\leq i\leq m}\ ,T_1\}$ 
		\STATE 
		$\mathbf{z}_k$: is the evaluation in the $k$th iteration\\
		$\mathbf{a}_i$: Gaussian vectors\\
		$y_i=|\langle\mathbf{a}_i,\mathbf{x}\rangle|^2$: measurements\\
		$\eta_i$: the parameter\\
		$T_1$: the maximum steps of gradient descent\\
		\ENSURE$\mathbf{z}_{k+1}$
		\vskip 4mm
		\hrule
		\vskip 2mm
		$\mathbf{Initialization}$
		\STATE
		Set $\mathbf{z}_{k+1}^0=\mathbf{z}_k$\\
		\vskip 2mm
		\FOR{$t=1$; $t\leq T_1$; $t++$ } 
		\STATE
		$\mathbf{z}_{k}^{t}=\mathbf{z}_{k}^{t-1}-\mu_t\nabla\mathit{f}^k(\mathbf{z}_{k}^t)$
		\ENDFOR\\
		$\mathbf{z}_{k+1}=\mathbf{z}_{k}^{T_1}$   	   	
	\end{algorithmic}  
\end{algorithm}
\indent
From algorithm 1, we can see that we will solve an optimization problem in each iteration $k$:
\begin{eqnarray}
\mathbf{z}_{k+1}\in\mathrm{argmin}~f^k(\mathbf{z})=\frac{1}{m}\sum_{i=1}^{m}f_i^k(\mathbf{z}),
\end{eqnarray}  
where $f_i^k(\mathbf{z})=\omega_i^k(|\langle\mathbf{a}_i,\mathbf{z}\rangle|^2-y_i)^2$.\\	\indent
For simplicity, we use gradient descent algorithm to deal with it in this paper. Details can be seen in algorithm 2. On the other hand, there are a wild range of alternatives which can also be used to solve (2.4). Sun et al. \cite{sun2016geometric} came up with the trust region method. Gao et al. \cite{gao2016gauss} utilized the Gauss-Newton method. Li et al. \cite{li2016gradient} suggested using the conjugate gradient method and LBFGS method.\\
\indent
It is critical to choose a proper stepsize $\mu_t$ in every step of gradient descent. In this paper, we use the backtracking method. For fairness, we also add this backtracking method into the WF and TWF for every numerical testing. The details of backtracking method can be seen in algorithm 3.\\
\begin{algorithm}
	\caption{\textbf{Stepsize Choosing via Backtracking Method}} 
	\label{alg:Framwork} %
	\renewcommand{\algorithmicrequire}{\textbf{Input:}}
	\renewcommand\algorithmicensure {\textbf{Output:} }
	\begin{algorithmic}   
		\REQUIRE$\{\mathit{f}(\mathbf{z}), \nabla\mathit{f}(\mathbf{z}),\mathbf{z}^{(k)},\beta\}$ 
		\STATE $\beta$ is a predetermined parameter which is in $(0,1)$
		\ENSURE$\mu^{(k)}$
		\vskip 4mm
		\hrule
		\vskip 2mm
		$\mathbf{General~step}$
	\end{algorithmic}  
	\begin{algorithmic}[1]   
		\STATE set $\tau=1$\\
		\STATE Repeat $\tau\leftarrow 0.5\tau$ until\\
		$\mathit{f}(\mathbf{z}^{(k)}-\tau\nabla\mathit{f}(\mathbf{z}^{(k)}))<\mathit{f}(\mathbf{z}^{(k)})-\tau\beta||\nabla\mathit{f}(\mathbf{z}^{(k)})||^2$
		\STATE $\mu^{(k)}=\tau$
	\end{algorithmic}  
\end{algorithm}
\indent
Empirically, the reweighted procedure will change the objective function $f^{k}$ which will prevent the algorithm from being stagnated into the local minimums all the time and proceed to search for the global optimum. When the ratio between $m$ and $n$ is comparatively large, the geometric property of the $\mathit{f}^{1}(\mathbf{z})$ is benign will fewer local minimums, then we can get a comparatively high accurate solution in several iterations. Figure 2.1 is the function landscape of $\mathit{f}^{1}(\mathbf{z})$ with $\mathbf{x}=\{[0.5;0.5],[-0.5,-0.5]\}$. We can see the weighted function is more steep than the unweighted one in the neighbor of the global optimums. From the geometrical prospect, the weighted function is alible to converge to the optimum. \\
\indent
Next, we will give the convergence analysis of RWF.
\subsection{Convergence of RWF}
\begin{figure}
	\centering
	\includegraphics[width=8cm,height=5cm]{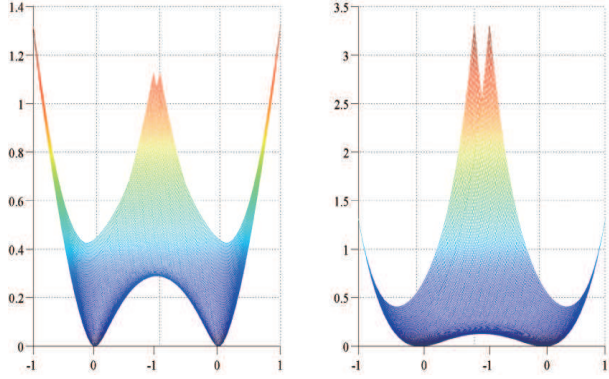}
	\caption{Left is the landscape of $f^1(\mathbf{z})$, right is the landscape of $f(\mathbf{z})$ defined by (1.2) . For both pictures, $\mathbf{x}=\big\{[0.5;0.5],[-0.5,-0.5]\big\}$, $m=100$.}
\end{figure}
\indent
To establish the convergence of RWF, firstly, we define the distance of any estimation $\mathbf{z}$ to the solution set as
\begin{eqnarray}
dist(\mathbf{z},\mathbf{x})
=\mathop {\mathrm{min}}\limits_{\phi\in[0,2\pi]}||\mathbf{z}-\mathbf{x}e^{\mathrm{j}\phi}||. 
\end{eqnarray} 
\indent
Lemma 2.1 and Lemma 2.2 give the bounds of $dist(\mathbf{z}_0,\mathbf{x})$. 
\begin{lemma}\cite{candes2015phase}
	Let $\mathbf{x}\in\mathbb{C}^n$ be an arbitrary vector, $\mathbf{y}=|\mathbf{A}\mathbf{x}|^2\in\mathbb{R}^m$ with $\mathbf{A}=[\mathbf{a}_1,...\mathbf{a}_m]^*$, $\mathbf{a}_i^*\overset{i.i.d}\sim\mathcal{N}(\mathbf{0},\mathbf{I})$. Then when $m\geq c_0nlogn$, where $c_0$ is a sufficiently large constant, the Wirtinger flow initial estimation $\mathbf{z}_0$ normalized to squared Euclidean norm equal to $m^{-1}\sum_iy_i$ obeys:
	\begin{eqnarray}
	dist(\mathbf{z}_0,\mathbf{x})\leq\frac{1}{8}||\mathbf{x}||.
	\end{eqnarray} 
\end{lemma}
\begin{lemma}\cite{candes2015phase}
	Let $\mathbf{x}\in\mathbb{C}^n$ be an arbitrary vector and assume collecting $L$ admissible coded diffraction pattern with $L\geq c_0(logn)^4$, where $c_0$ is a sufficiently large numerical constant. Then $\mathbf{z}_0$ satisfies:
	\begin{eqnarray}
	dist(\mathbf{z}_0,\mathbf{x})\leq\frac{1}{8\sqrt{n}}||\mathbf{x}||.
	\end{eqnarray}
\end{lemma}
\indent
The selection of $\eta_i$ did have effect on the performance of RWF. As Figure 2.2 depicted, different $\alpha$ can have different convergence rate for RWF.
\begin{figure}
	\centering
	\includegraphics[width=7cm,height=6cm]{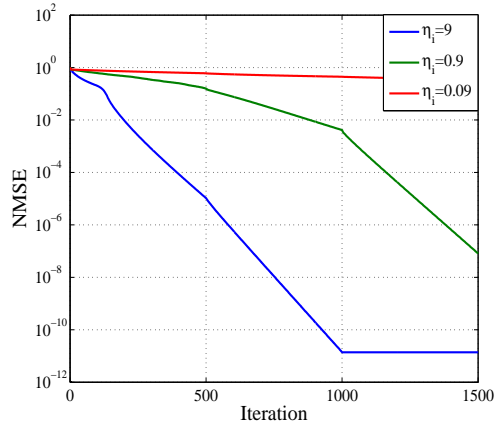}
	\caption{The $\mathbf{NMSE}$ of different $\alpha$ for RWF. $m/n=3$ for $\mathbf{x}\in \mathrm{R}^{256}$}
\end{figure}
For convenience to prove, we assume $\eta_i=0.9$ to be a constant for all $i$. Utilizing the Lemmas above, we can establish the convergence theory of the RWF.
\
\begin{theorem}
	Let $\mathbf{z}_k$ be the evaluation of the $k$th iteration in algorithm 1. If 
	$\mathop{\mathrm{max}}\limits_{i=1,...,m}\Big|\big|\langle\mathbf{a}_i,\mathbf{z}_k\rangle\big|^2-y_i\Big|<0.1$. Namely, $1\leq\omega_i\leq\frac{10}{9}$. Then taking a constant stepsize $\mu_t=\mu$ for $t=1,2,...$ with $\mu<c_1/n$ for some fixed numerical $c_1$. Then with probability at least $1-13e^{-\lambda n}-me^{-1.5m}-8/{n^2}$ for some constant term $\lambda$, the estimation in algorithm 1 satisfying:
	\begin{eqnarray}
	dist(\mathbf{z}_k^t,\mathbf{x})\leq\frac{1}{8}(1-\frac{\mu}{4})^{t-1}||\mathbf{x}||.
	\end{eqnarray}
\end{theorem}
The details of proof are given in appendix.\\
\indent
Note that $\mathbf{x}$ is the global optimum of (2.4) for each $k$. The solution of (2.4) can get close enough to $\mathbf{x}$ during the iterations. we assume $dist(\mathbf{z}_k,\mathbf{x})\leq dist (\mathbf{z}_0,\mathbf{x})$ in the $kth$ iteration. During the iteration, $\mathbf{z}_k$  moves to the region $E(\mathbf{z})$, where
\begin{eqnarray*}
	E(\mathbf{z})=\Big\{\mathbf{z}\Big|\mathop {\mathrm{max}}\limits_{ i=1,...,m}\Big|\big|\langle\mathbf{a}_i,\mathbf{z}\rangle\big|^2-y_i\Big|<0.1\Big\}.
\end{eqnarray*}
Once we find $\mathbf{z}_k$ dropping into $E(\mathbf{z})$, there can be a geometric convergence of RWF~by theorem 2.3. Theorem 2.3 also ensures the exponential convergence of WF becauese it is a special case of RWF. Figure 2.2 shows how RWF converges to the optimum. Here, the maximum steps of gradient descent in RWF during each iteration is $500$. $m/n=2.5$, $n=256$. At first 500 steps, both algorithm can't let $\mathbf{NMSE}$(The definition of $\mathbf{NMSE}$ can be found in section 3) decrease and get stuck into the local minimums. The reweighted procedure can empirically pull out the evaluation from these holes by changing the objective function. So RWF can continue to search for the optimum.\\
\begin{figure}
	\centering
	\includegraphics[width=7cm,height=6cm]{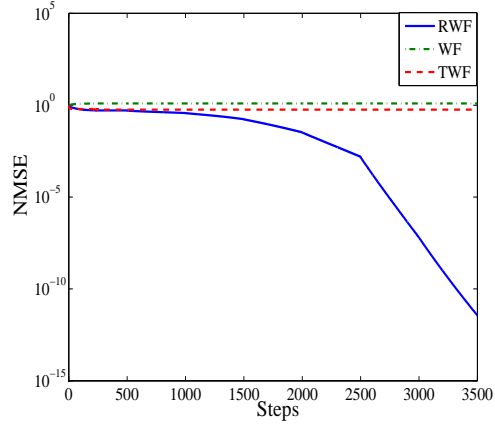}
	\caption{The $\mathbf{NMSE}$ of different methods during iteration. $m/n=2.5$ for $\mathbf{x}\in \mathrm{R}^{256}$}
\end{figure}
\section{Numerical testing}
Numerical results are given in this section which show the performance of RWF together with WF and TWF. All the tests are carried out on the Lenovo desktop with a 3.60 GHz Intel Corel i7 processor and 4GB DDR3 memory. Here,  we are in favor of the normalized mean square error($\mathbf{NMSE}$) which can be calculated as below:
\begin{eqnarray*}
	\centering
	\mathbf{NMSE}=\frac{\mathrm{dist}(\hat{\mathbf{x}},\mathbf{x})}{||\mathbf{x}||},
\end{eqnarray*}
where $\hat{\mathbf{x}}$ is the numerical estimation of $\mathbf{x}$.\\
\indent
In the simulation, $\mathbf{x}$ was created by real Gaussian random vector $\mathcal{N}(\mathbf{0},\mathbf{I})$ or the complex Gaussian random vector $\mathcal{N}(\mathbf{0},\mathbf{I}/2)+j\mathcal{N}(\mathbf{0},\mathbf{I}/2)$. $\{\mathbf{a}_i\}_{1\leq i\leq m}$ are drawn i.i.d from either  $\mathcal{N}(\mathbf{0},\mathbf{I})$ or $\mathcal{N}(\mathbf{0},\mathbf{I}/2)+j\mathcal{N}(\mathbf{0},\mathbf{I}/2)$. In all simulating testings, $\eta_i=0.9$ to be a constant. Figure 3.1 shows the exact recovery rate of RWF, TWF and WF. The length $n$ of $\mathbf{x}$ is $256$. For RWF, we set the maximum iteration to be 300 and the maximum steps of gradient desent in each iteration to be 500. For fairness, the maximum iterations of WF and TWF are both 150000. Once the $\mathbf{NMSE}$ is less than $10^{-5}$, we stop iteration and  judge the real signal is exactly recovered. Let $m/n$ vary from 1 to 8, at each $m/n$ ratio, we do $50$ replications. The empirical recovery rate is the total successful times divides 50 at each $m/n$.\\
\indent
The performance of different algorithms in the simulation testing discussed above is shown in Figure 3.1. In figure 3.1(a), for real-valued signal, RWF can have a 90\% successful recover rate of the signals when  $m=2.4n$. There are even some instances of successful recovery when $m=1.6n$. In contrast, TWF and RWF need $m=3.3n$ and $m=4.6n$ to recover the signals with the same recovery rate as RWF. In figure 3.1(b), for the complex case, the performance of RWF is superior than TWF and WF too. RWF can nearly get a high recovery probability about 85\% at $m=3.5n$. The sampling complexity of RWF is approximately to the information-limits from those numerical simulations which demonstrates the high capability of RWF to deal with the PR problem. \\
\begin{figure}	
	\centering
	\begin{subfigure}[t]{3in}
		\centering
		\includegraphics[width=3in]{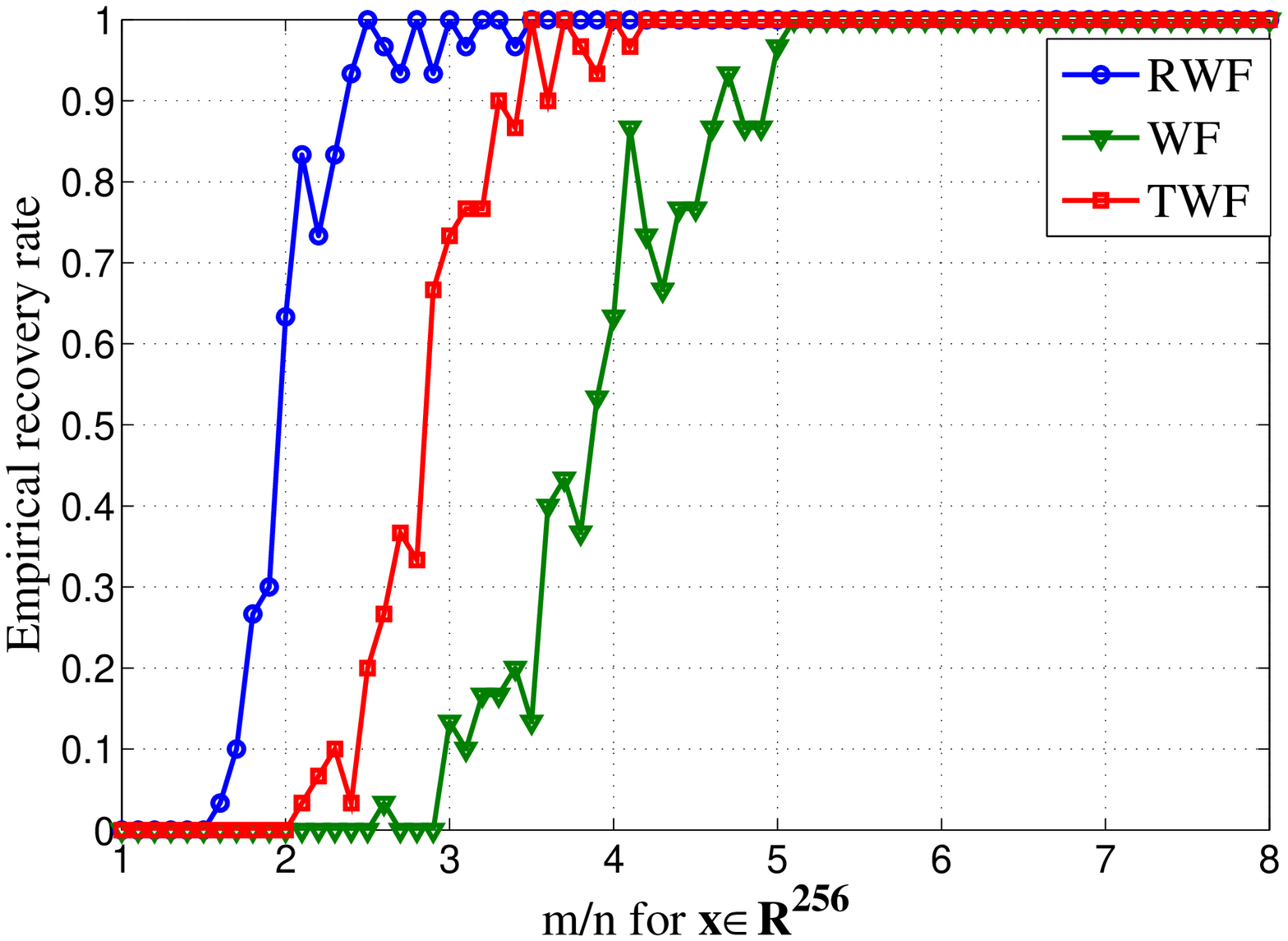}
		\caption{}\label{fig:1a}		
	\end{subfigure}
	\quad
	\begin{subfigure}[t]{3in}
		\centering
		\includegraphics[width=3in]{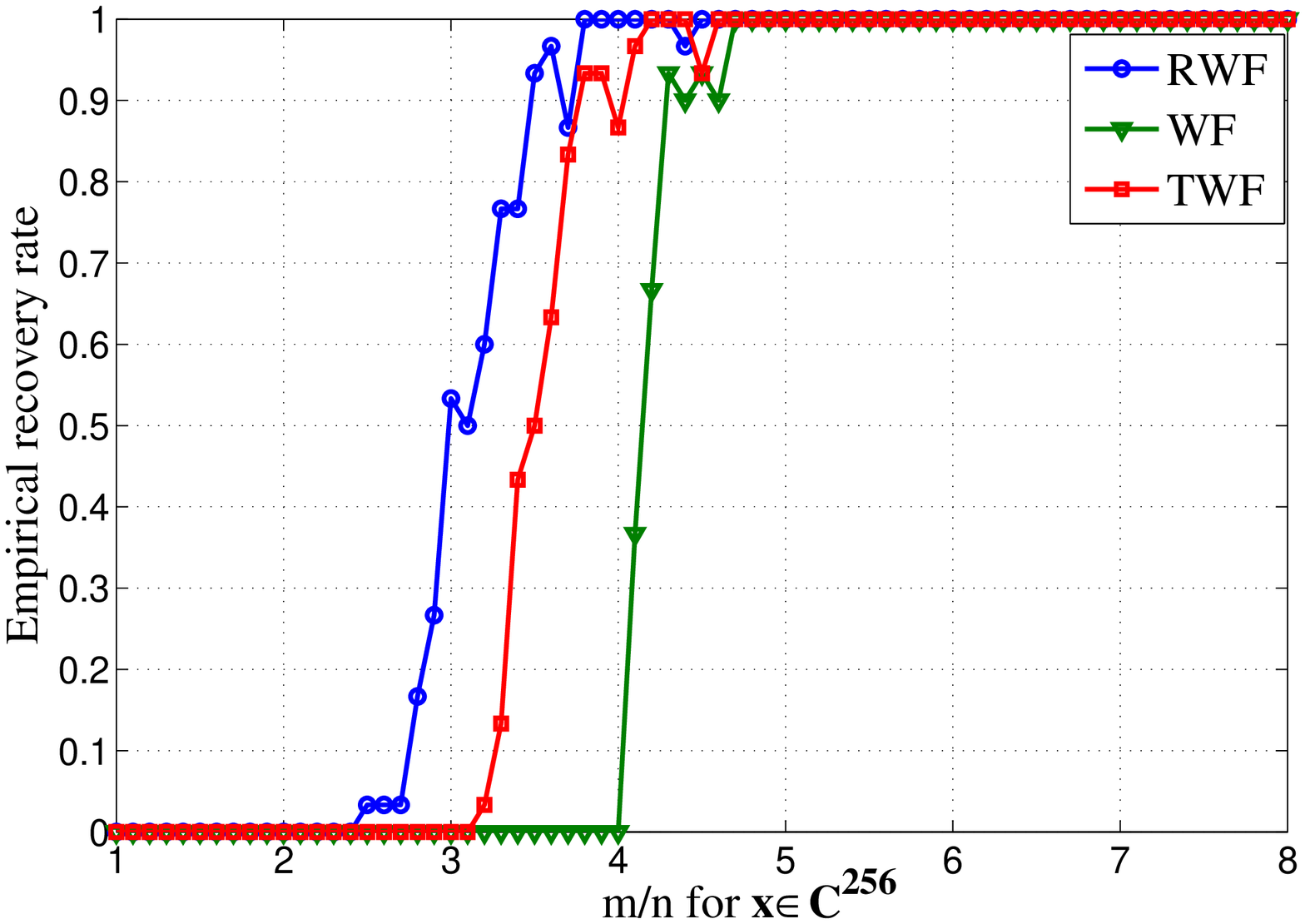}
		\caption{}\label{fig:1b}
	\end{subfigure}
	\caption{The comparison between RWF, WF and TWF for the noiseless signal.}\label{fig:1}
\end{figure}
\indent
The times of iteration determine the convergence rate of RWF. Figure 3.2 shows the times of iteration need for RWF to be recover the signal where $m/n$ is from $2$ to $8$. At each $m/n$, we record 50 times successful tests and average their iteration times. The maximum steps of gradient descent in each iteration are 500. If the $\mathbf{NMSE}$ is below $10^{-5}$, we declare this trial successful. We can see that when the $m/n$ is small, RWF need more iterations to search for the global minimum, this procedure need more computation costs. With $m/n$ increasing, the iterations gradually decrease which is nearly equal to one because of the benign geometric property of objective function when $m/n$ becomes large.\\  
\begin{figure}	
	\centering
	\begin{subfigure}[t]{3in}
		\centering
		\includegraphics[width=3in]{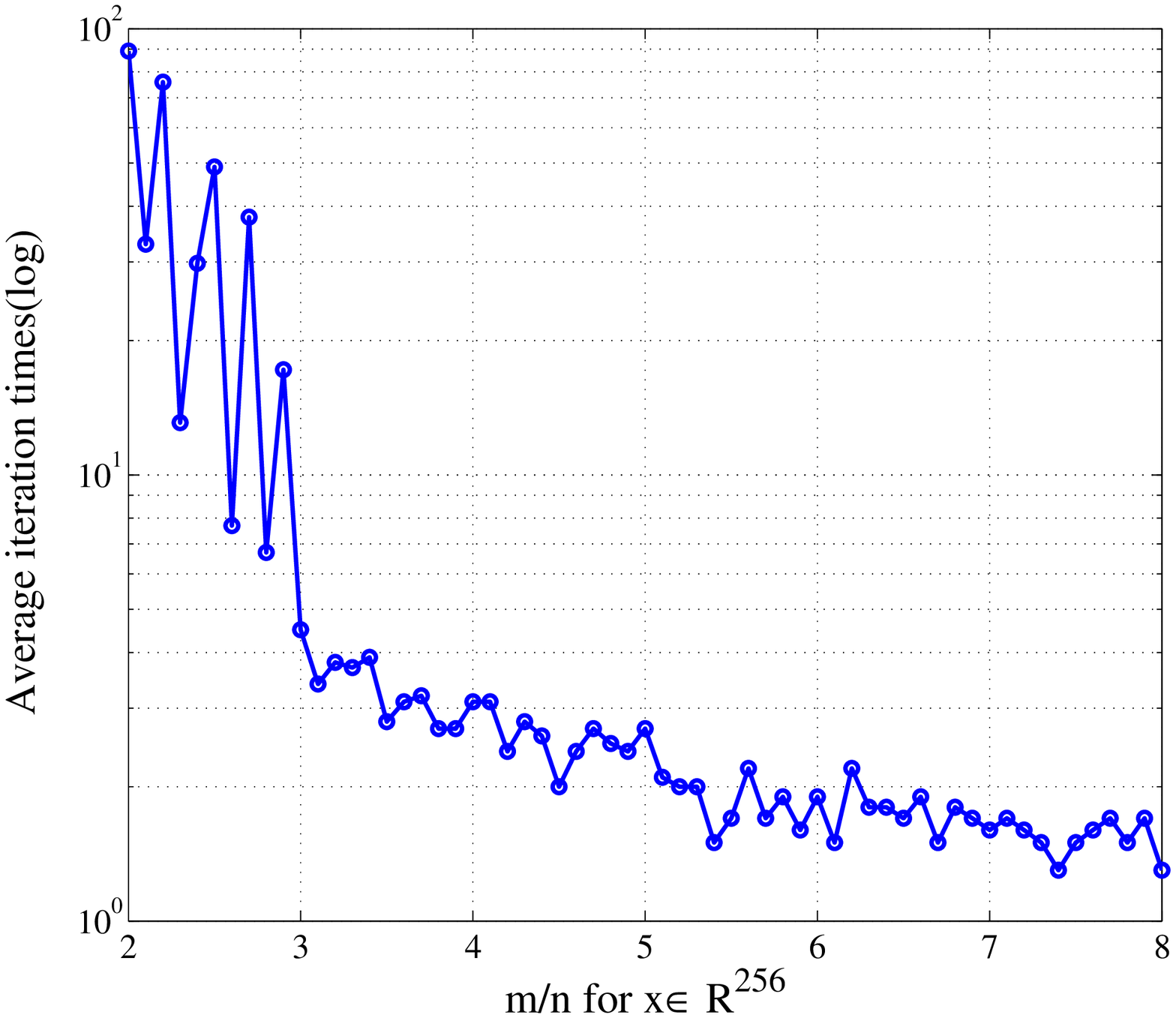}
		\caption{}\label{fig:1a}		
	\end{subfigure}
	\quad
	\begin{subfigure}[t]{3in}
		\centering
		\includegraphics[width=3in]{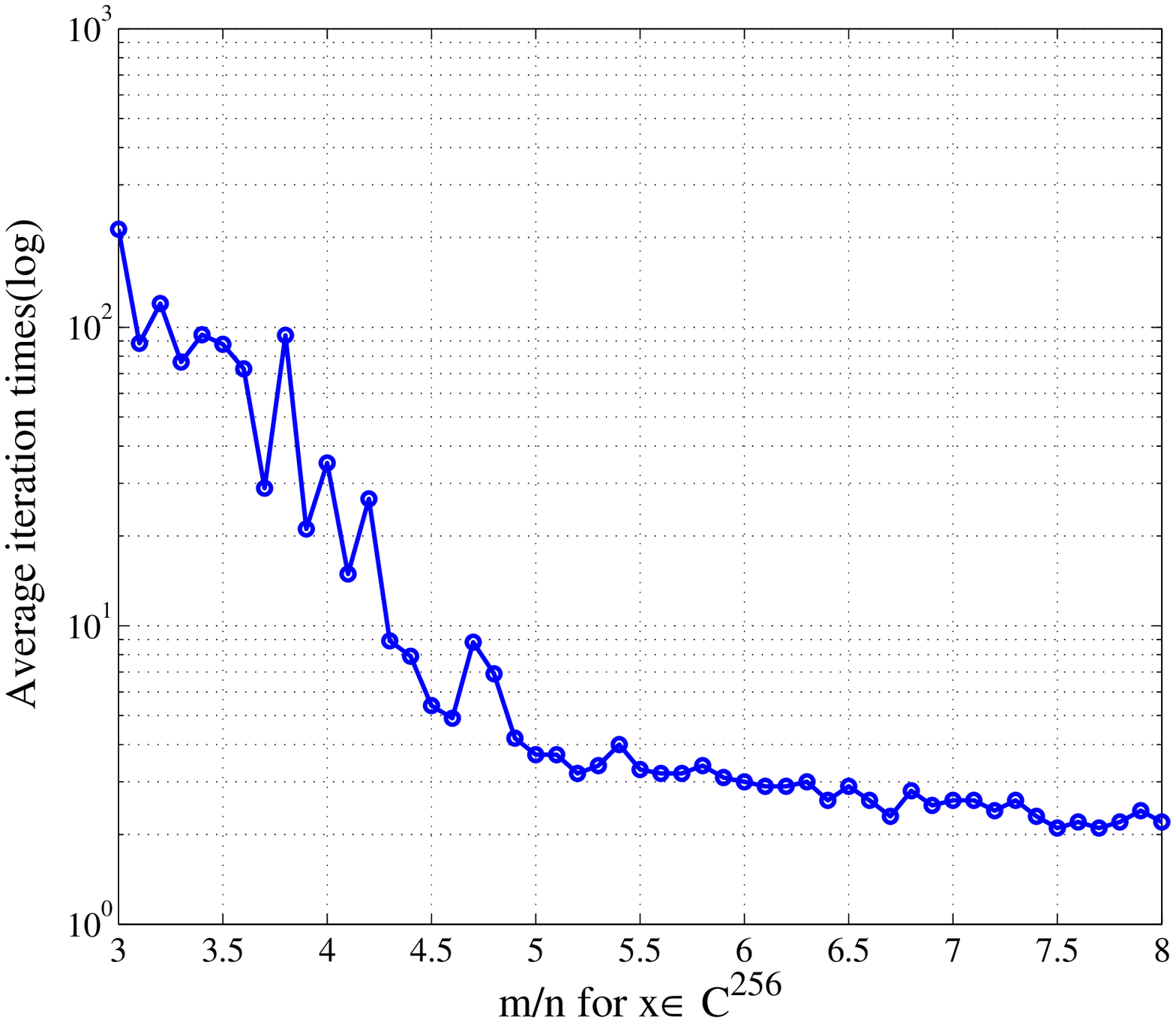}
		\caption{}\label{fig:1b}
	\end{subfigure}
	\caption{The average iteration times of RWF.}\label{fig:1}
\end{figure}
\indent Our reweighted idea can also be extended to the Code diffraction pattern(CDP). The details of CDP model can be referred in \cite{Cand2013Phase}. The weighted CDP model can be described as below:\\
\begin{eqnarray}
y_r^k=\big|\sum_{t=0}^{n-1}\omega_r^kz^k[t]\bar{d}_l(t)e^{-j2\pi k_1t/n}\big|^2,\\
r=(k_1,l),0\leqslant k_1\leqslant n-1,1\leqslant l\leqslant L\nonumber,
\end{eqnarray}
where
\begin{eqnarray*}
	\omega^k_r=1\Big/\Big(\Big|\big|\sum_{t=0}^{n-1}z^{k-1}[t]\bar{d}_l(t)e^{-j2\pi k_1t/n}\big|^2-y_r\Big|+\eta_r\Big),
\end{eqnarray*}
\begin{eqnarray*}
	y_r=\big|\sum_{t=0}^{n-1}x[t]\bar{d}_l(t)e^{-j2\pi k_1t/n}\big|^2,\\
	r=(k_1,l),0\leqslant k_1\leqslant n-1,1\leqslant l\leqslant L,
\end{eqnarray*} 
where $\mathbf{x}$ is the real signal. We can also get the high accuracy evaluation of $\mathbf{x}$ with algorithm 1. Figure 3.3 is the comparison between RWF and WF with CDP model. We generated $\mathbf{x}$ from $ \mathcal{N}(0,\mathbf{I}/2)+j\mathcal{N}(0,\mathbf{I}/2)$. The length of $\mathbf{x}$ is 256. L varies from 2 to 8. Other settings are the same as those for real or complex Gaussian signal above. We can see that RWF has a little advantage over WF for 1D CDP model.\\
\begin{figure}
	\centering
	\includegraphics[width=7cm,height=6cm]{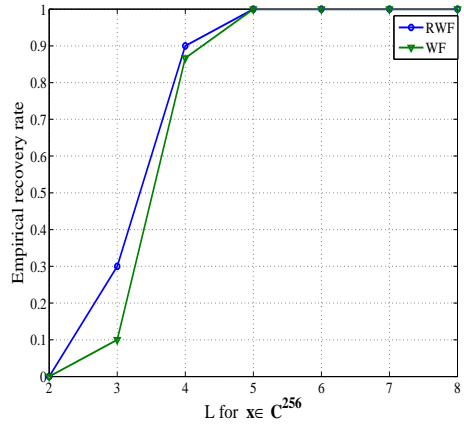}
	\caption{The comparison between RWF and WF with CDP model for noiseless signal}
\end{figure} 
\indent
Figure 3.4 is the result of 2D CDP model. These are 3D Caffein molecules projected on the 2D plane. The size of picture is 128$\times$128. L is 7. Because it is a RGB picture. As a result, we apply RWF and WF for every R,G,B channels independly. For RWF, we set the maximum iteration to be 300 and the maximum steps of gradient descent in each iteration to be 500. The maximum iteration is 150000 for WF. We can see that the picture recovered by RWF is better than that recovered by WF. 
\begin{figure}	
	\centering
	\begin{subfigure}[t]{1.5in}
		\centering
		\includegraphics[width=1.5in]{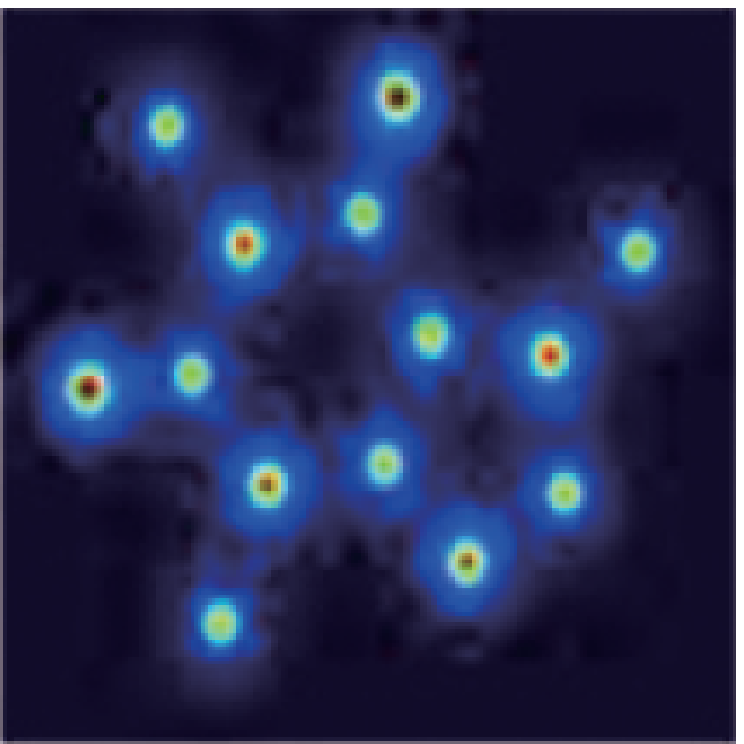}
		\caption{}\label{fig:1a}		
	\end{subfigure}
	\quad
	\begin{subfigure}[t]{1.5in}
		\centering
		\includegraphics[width=1.5in]{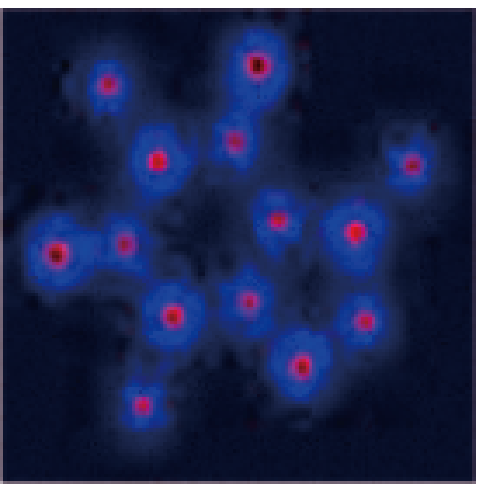}
		\caption{}\label{fig:1b}
	\end{subfigure}
	\quad
	\begin{subfigure}[t]{1.5in}
		\centering
		\includegraphics[width=1.5in]{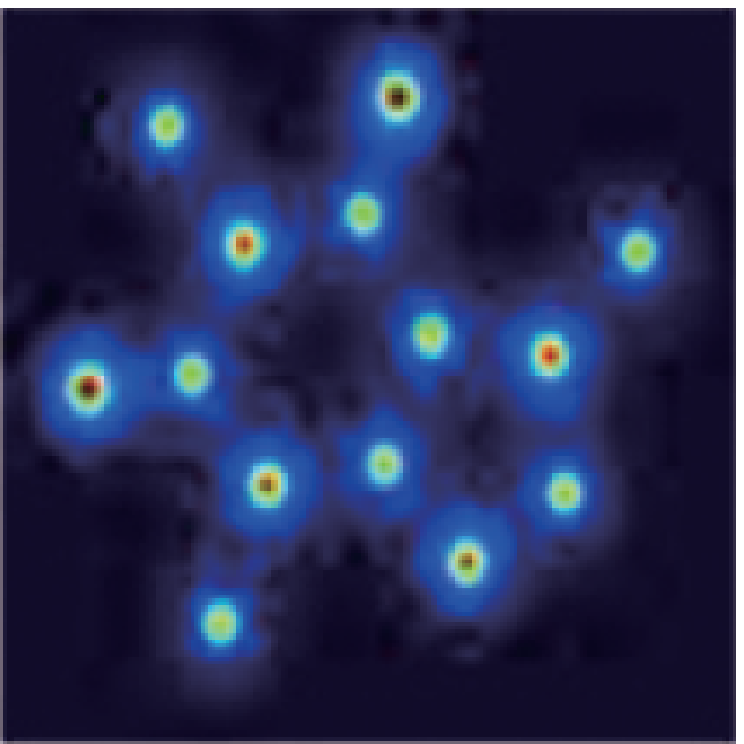}
		\caption{}\label{fig:1b}
	\end{subfigure}
	\caption{The 3D Caffein molecule. (a) is the real, (b) is recovered by WF, (c) is recovered by RWF .}\label{fig:1}
\end{figure}
\section{Conclusion}
In this paper, we propose a reweighted Wirtinger flow algorithm for phase retrieval problem. It can make the gradient descent algorithm more alible to converge to the global minimum when the sampling complexity is low by reweighting the objective function in each iteration. But comparing to WF, this algorithm has more computational cost. So in the future work, we will be keen to accelerate it such as using stochastic gradient method. 
\section{Acknowledgement}
This work was supported in part by National Natural Science foundation(China): 0901020115014.



\bibliography{1.bib}
\section{Appendix}
\subsection{Preliminaries}
For $a_1\sim\mathcal{N}(0,1)$, $a_2\sim\mathcal{CN}(0,1)$, we can have the equality below:
\begin{eqnarray*}
	&&\mathbb{E}(|a_1|^{2p})=(2p-1)!!,\\
	&&\mathbb{E}(|a_2|^{2p})=p!,
\end{eqnarray*}
where $p\in\mathbb{Z}^+$.
As discussed in \cite{candes2015phase}, measurement vector $\mathbf{a}_i$ for $i=1,...,m$ are supposed to satisfy the inequality $||\mathbf{a}_i||\leq\sqrt{6n}$ in the Gaussian model with probaility at least $1-me^{-1.5n}$. In the CDP model with admissible CDPs, the inequality $||\mathbf{a}_i||\leq\sqrt{6n}$ for $i=1,...,m$ also holds. Next, we will introduce the RC condition. 
\begin{definition}(Regularity Condition)
	The function $f$ is called to satisfy the regularity condition($\mathrm{RC}$) if:
	\begin{eqnarray*}
		&~&\mathrm{Re}(\langle\nabla f(\mathbf{z}),\mathbf{z}-\mathbf{x}e^{j\phi(\mathbf{z})}\rangle)\\&\geqslant& \frac{1}{\alpha}dist^2(\mathbf{z},\mathbf{x})
		+\frac{1}{\beta}||\nabla f(\mathbf{z})||^2,
	\end{eqnarray*}
\end{definition}
holds for all $\mathbf{z}\in E(\mathbf{z})$. where $E(\mathbf{z})$ is the region where $\Big\{\mathbf{z}\Big|\mathop {\mathrm{max}}\limits_{ i=1,...,m}\Big|\big|\langle\mathbf{a}_i,\mathbf{z}\rangle\big|^2-y_i\Big|<0.1\Big\}$.\\
\indent
Followed by the results in\cite{candes2015phase}, we will prove that geometric convergence can be guaranteed if $f$ satisfy the RC condition.
\begin{eqnarray*}
	dist^2(\mathbf{z}-\mu\nabla f(\mathbf{z}),\mathbf{x})\leq(1-\frac{2\mu}{\alpha})dist^2(\mathbf{z},\mathbf{x}).
\end{eqnarray*}
\subsection{Proof of Regularity condition}
The proof is followed by \cite{candes2015phase} which proves the gradient satisfying the local smoothness and local curvature conditions. Then we will combine the inequalities in two those conditions to proof regularity condition. First, we will introduce Local Curvature Condition and Local Smoothness Condition.
\begin{definition}(Local Curvature Condition):
	The function f satisfies the local curvature condition for all $\mathbf{z}\in E(\mathbf{z})$,
	\begin{eqnarray*}
		&&\mathrm{Re}(\textless\nabla f(\mathbf{z}),\mathbf{z}-\mathbf{x}e^{j\phi(\mathbf{z})}\textgreater)\\&\geqslant& (\frac{1}{\alpha}+\frac{1+\delta}{4})dist^2(\mathbf{z},\mathbf{x})+\frac{1}{10m}\sum_{i=1}^{m}|\mathbf{a}_i^*(\mathbf{z}-e^{i\phi(\mathbf{z})}\mathbf{x})|^4.
	\end{eqnarray*}
\end{definition}
\begin{definition}(Local Smoothness Condition):
	The function $f$ satisfies the local smoothness condition, in fact for all 
	vectors $\mathbf{z}\in E(\mathbf{z})$ we have:
	\begin{eqnarray}
	||\nabla f(\mathbf{z})||^2&\leq&\beta\Big((\frac{1+\delta}{4})dist^2(\mathbf{z},\mathbf{x})\nonumber\\
	&~&+\frac{1}{10m}\sum_{i=1}^{m}|\mathbf{a}_i^*(\mathbf{z}-e^{j\phi(\mathbf{z})}\mathbf{x})|^4\Big).
	\end{eqnarray}
\end{definition}
Local Curvature Condition implies that the function curves sufficiently upwards in the neighborhood of the global minimizers. Local Smoothness Condition shows that the gradient of the function doesn't vary too much near the curves of the global optimizers. Next, we will prove those two conditions respectively
\subsection{Proof of the Local Curvature Condition}
For any $\mathbf{z}\in E(\mathbf{z})$, we want to prove Local Curvature Condition. Recall that
\begin{eqnarray*}
	\nabla f(\mathbf{z})=\frac{1}{m}\sum_{i=1}^{m}\omega_i(|\mathbf{a}_i^*\mathbf{z}|^2-y_i)^2(\mathbf{a}_i\mathbf{a}_i^*)\mathbf{z}.
\end{eqnarray*}
In this paper, $\eta$ is chosen as a constant number $0.9$. So in the region $E(\mathbf{z})$, $1\leq\omega_i\leq\frac{10}{9}$. We define $\mathbf{h}:=e^{-j\phi(z)}\mathbf{z}-\mathbf{x}$. What we shall prove is:
\begin{eqnarray}
&&\frac{1}{m}\sum_{i=1}^{m}(2\omega_i\mathrm{Re}(\mathbf{h}^*\mathbf{a}_i\mathbf{a}_i^*\mathbf{x})^2\nonumber\\
&&+3\omega_i\mathrm{Re}(\mathbf{h}^*\mathbf{a}_i\mathbf{a}_i^*\mathbf{x})|\mathbf{a}_i^*\mathbf{h}|^2+\omega_i|\mathbf{a}_i^*\mathbf{h}|^4)\nonumber\\
&\geq&(\frac{1}{\alpha}+\frac{1+\delta}{4})||\mathbf{h}||^2+\frac{1}{10m}\sum_{i=1}^{m}|\mathbf{a}_i^*\mathbf{h}|^4,
\end{eqnarray} 
holds for all $\mathbf{h}$, with $\mathrm{Im}(\mathbf{h}^*\mathbf{x})=0$, $||\mathbf{h}||\leq\epsilon$. From (2.5),(2.6) we can assume: $\epsilon=\frac{1}{8}||\mathbf{x}||$ for Gaussian model, $\epsilon=\frac{1}{8\sqrt{n}}||\mathbf{x}||$ for the CDP model. The assuming here is reasonable, because in each step we will solve an optimization problem whose optimum is $\mathbf{x}$, $\mathbf{h}$ constantly decreases during the iteration, so the inequalities also hold when $\mathbf{z}$ is in $E(\mathbf{z})$. Equivalently, we only need to prove that for all $\mathbf{h}$ satisfying $\mathrm{Im}(\mathbf{h}^*\mathbf{x})=0$, $||\mathbf{h}||_2=1$ and for all $s$ with $0\leq s\leq \epsilon$,
\begin{eqnarray}
&&\frac{1}{m}\sum_{i=1}^{m}\Big(2\omega_i\mathrm{Re}(\mathbf{h}^*\mathbf{a}_i\mathbf{a}_i^*\mathbf{x})^2+3\omega_rs\mathrm{Re}(\mathbf{h}^*\mathbf{a}_i\mathbf{a}_i^*\mathbf{x})|\mathbf{a}_i^*\mathbf{h}|^2\nonumber\\
&~&+(\omega_i-\frac{1}{10})s^2|\mathbf{a}_i^*\mathbf{h}|^4\Big)\nonumber\\
&~~~&\geq\frac{1}{\alpha}+\frac{1+\delta}{4}
\end{eqnarray}
By Corollary 7.5 in \cite{candes2015phase}, with high probability,
\begin{eqnarray}
\frac{1}{m}\sum_{r=1}^{m}\mathrm{Re}(\mathbf{h}^*\mathbf{a}_r\mathbf{a}_r^*\mathbf{x})^2\geq\frac{1-\delta}{2}+\frac{3}{2}\mathrm{Re}(\mathbf{x}^*\mathbf{h})^2
\end{eqnarray}
holds for all $\mathbf{h}$ with length $||\mathbf{h}||=1$. To prove (5.2), it is sufficient to prove:
\begin{eqnarray}
&&\frac{1}{m}\sum_{i=1}^{m}\Big((\omega_i+\frac{1}{2})\omega_i\mathrm{Re}(\mathbf{h}^*\mathbf{a}_i\mathbf{a}_i^*\mathbf{x})^2\nonumber\\
&~&+3\omega_is\mathrm{Re}(\mathbf{h}^*\mathbf{a}_i\mathbf{a}_i^*\mathbf{x})|\mathbf{a}_i^*\mathbf{h}|^2+(\omega_i-\frac{1}{10})s^2|\mathbf{a}_i^*\mathbf{h}|^4\Big)\nonumber\\
&~~~&\geq(\frac{1}{\alpha}+\frac{1}{2})+\frac{3}{4}\mathrm{Re}(\mathbf{x}^*\mathbf{h})^2
\end{eqnarray}
\indent
1) Proof of (6.5). With $\epsilon=\frac{1}{8}\sqrt{n}$ in the Gaussian and CDP models: set $\epsilon=\frac{1}{8}\sqrt{n}$. We show that with high probability, (6.5) holds for all $\mathbf{h}$ satisfying $\mathrm{Im}(\mathbf{h}^*\mathbf{x})=0$, $||\mathbf{h}||_2=1, 0\leq s\leq\epsilon,\delta\leq\frac{23}{601},\alpha\geq10$. By Cauchy-Schwarz inequality:
\begin{eqnarray}
&~&\frac{1}{m}\sum_{i=1}^{m}\Big((\omega_i+\frac{1}{2})\mathrm{Re}(\mathbf{h}^*\mathbf{a}_i\mathbf{a}_i^*\mathbf{x})^2\nonumber\\
&&+3\omega_rs\mathrm{Re}(\mathbf{h}^*\mathbf{a}_i\mathbf{a}_i^*\mathbf{x})|\mathbf{a}_i^*\mathbf{h}|^2+(\omega_i-\frac{1}{10})s^2|\mathbf{a}_i^*\mathbf{h}|^4\Big)\nonumber\\
&\geq&\frac{5}{2m}\sum_{i=1}^{m}\mathrm{Re}(\mathbf{h}^*\mathbf{a}_i\mathbf{a}_i^*\mathbf{x})^2\nonumber\\
&&-\frac{10s}{3}\sqrt{\frac{1}{m}\sum_{i=1}^{m}\mathrm{Re}(\mathbf{h}^*\mathbf{a}_i\mathbf{a}_i^*\mathbf{x})^2}\sqrt{\frac{1}{m}\sum_{i=1}^{m}|\mathbf{a}_i^*\mathbf{h}|^4}\nonumber\\
&&+\frac{9s^2}{10}\frac{1}{m}\sum_{i=1}^{m}|\mathbf{a}_i^*\mathbf{h}|^4\nonumber\\
&&=\Big(\sqrt{\frac{5}{2}}\sqrt{\frac{1}{m}\sum_{i=1}^{m}\mathrm{Re}(\mathbf{h}^*\mathbf{a}_i\mathbf{a}_i^*\mathbf{x})^2}\nonumber\\
&&-\frac{5s}{3}\sqrt{\frac{2}{5m}\sum_{i=1}^{m}|\mathbf{a}_i^*\mathbf{h}|^4}\Big)^2+(-\frac{19}{90})s^2\frac{1}{m}\sum_{i=1}^{m}|\mathbf{a}_i^*\mathbf{h}|^4\nonumber\\
&\geq&\frac{5}{4m}\sum_{i=1}^{m}\mathrm{Re}(\mathbf{h}\mathbf{a}_i\mathbf{a}_i^*\mathbf{x})^2\nonumber\\
&&+(-\frac{19}{90})s^2\frac{1}{m}\sum_{i=1}^{m}|\mathbf{a}_i^*\mathbf{h}|^4
\end{eqnarray}
The last equality follows from $(a-b)^2\geq\frac{a^2}{2}-b^2$.\\
By applying Lemma 7.8 in \cite{candes2015phase},
\begin{eqnarray}
\frac{1}{m}\sum_{i=1}^{m}|\mathbf{a}_i^*\mathbf{h}|^4&\leq&(\mathop {\mathrm{max}}\limits_{i}||\mathbf{a}_i||^2)(\frac{1}{m}\sum_{i=1}^{m}|\mathbf{a}_i^*\mathbf{h}|^2)\nonumber\\
&\leq&6(1+\delta)n
\end{eqnarray}
holds with high probability, Plugging (6.4) and (6.7) in (6.6) we will have:
\begin{eqnarray*}
	&~&\frac{1}{m}\sum_{i=1}^{m}\big((\omega_i+\frac{1}{2})\omega_i\mathrm{Re}(\mathbf{h}^*\mathbf{a}_i\mathbf{a}_i^*\mathbf{x})^2\\
	&&+3\omega_is\mathrm{Re}(\mathbf{h}^*\mathbf{a}_i\mathbf{a}_i^*\mathbf{x})|\mathbf{a}_i^*\mathbf{h}|^2+(\omega_i-\frac{1}{10})s^2|\mathbf{a}_i^*\mathbf{h}|^4\big)\\
	&\geq&\frac{3}{2}\mathrm{Re}(\mathbf{x}^*\mathbf{h})^2+\frac{599}{960}-\frac{601}{900}\delta\\
	&\geq&\frac{3}{2}\mathrm{Re}(\mathbf{x}^*\mathbf{h})^2+\frac{1}{2}+\frac{1}{10}\\
\end{eqnarray*}
Which follows by $\delta\leq\frac{23}{601}$ and $\alpha\geq10$.\\
\indent
2) Proof of (6.5) with $\epsilon=\frac{1}{8}$ in the Gaussian Model. We also utilize the skills in \cite{candes2015phase}. 
\begin{lemma}
	\cite{Bentkus2003An} Supposing $X_1,X_2,...,X_m$ are i.i.d real-valued random variables obeying $X_r\leq b$ for some nonrandom $b>0$, $\mathbb{E}X_r=0$ and $\mathbb{E}X_i^2=v^2$. Setting $\delta^2=m~max(b^2,v^2)$.
	\begin{eqnarray*}
		P(X_1+...+X_m\geq y)\leq min\Big(e^{-\frac{y^2}{2\delta^2}},c_0(1-\phi(y/\delta)\Big)
	\end{eqnarray*}
\end{lemma}
Where one can take $c_0=25$.\\
To prove this, we first prove it for a fixed $\mathbf{h}$, and then using a covering argument. \\
Define:
\begin{eqnarray*}
	Y_i:=Y_i(\mathbf{h},s)&=&(\omega_r+\frac{1}{2})\mathrm{Re}(\mathbf{h}^*\mathbf{a}_i\mathbf{a}_i^*\mathbf{x})^2\\
	&&+3\omega_is\mathrm{Re}(\mathbf{h}^*\mathbf{a}_i\mathbf{a}_i^*\mathbf{x})|\mathbf{a}_i^*\mathbf{h}|^2\\
	&&+(\omega_i-\frac{1}{10})s^2|\mathbf{a}_i^*\mathbf{h}|^4
\end{eqnarray*}
By lemma 7.3 in \cite{candes2015phase}:
\begin{eqnarray}
\mathbb{E}[\mathrm{Re}(\mathbf{h}^*\mathbf{a}_i\mathbf{a}_i^*\mathbf{x})^2]=\frac{1}{2}+\frac{3}{2}(\mathrm{Re}(\mathbf{x}^*\mathbf{h}))^2
\end{eqnarray}
and
\begin{eqnarray}
\mathbb{E}[\mathrm{Re}(\mathbf{h}^*\mathbf{a}_i\mathbf{a}_i^*\mathbf{x})^2|\mathbf{a}_i^*\mathbf{h}|^2]=2\mathrm{Re}(\mathbf{h}^*\mathbf{x})
\end{eqnarray}
$\mu_i=\mathbb{E}Y_i=(\omega_i+\frac{1}{2})\big(\frac{1}{2}+\frac{3}{2}\mathrm{Re}(\mathbf{x}^*\mathbf{h})^2\big)+6s\omega_i\mathrm{Re}(\mathbf{x}^*\mathbf{h})+3(\omega_i-\frac{1}{10})s^2$\\
using $s\leq\frac{1}{8}$ and $\omega_i\leq\frac{10}{9}$, we can conclude $\mu_i\leq7$\\
Because
\begin{eqnarray*}
	Y_i(h,s)&=&(3s\omega_i\mathrm{Re}(\mathbf{h}^*\mathbf{a}_i\mathbf{a}_i^*\mathbf{x})+\frac{1}{2}|\mathbf{a}_i^*\mathbf{h}|^2)^2\\
	&&+(\omega_i+\frac{1}{2}-9s^2\omega_i^2)\mathrm{Re}(\mathbf{h}^*\mathbf{a}_i\mathbf{a}_i^*\mathbf{x})^2\\
	&&+(\omega_i-\frac{1}{10}-\frac{1}{4})|\mathbf{a}_i^*\mathbf{h}|^4\\ 
	&\geq&0
\end{eqnarray*}
As a result, defining $X_i=\mu_i-Y_i$, so $X_i\leq\mu_i<6$. We can bound $\mathbb{E}X_i^2$ using (6.8) (6.9) and Holder inequality with $s\leq\frac{1}{8}$\\
We can have $\mathbb{E}X_i^2\leq\mathbb{E}Y_i^2$, besides:
\begin{eqnarray*}
	\mathbb{E}Y_i^2&=&(2\omega_i+\frac{1}{2})^2\mathbb{E}[\mathrm{Re}(\mathbf{h}^*\mathbf{a}_i\mathbf{a}_i^*\mathbf{x})^4]\\
	&&+(\omega_i-\frac{1}{10})^2s^4\mathbb{E}[|\mathbf{a}_i^*\mathbf{h}|^8]\\
	&&+6s\omega_i(2\omega_i+\frac{1}{2})\mathbb{E}(\mathrm{Re}(\mathbf{h}^*\mathbf{a}_i\mathbf{a}_i^*\mathbf{x})^3|\mathbf{a}_i^*\mathbf{h}|^2)\\
	&&+6s\omega_i(\omega_i-\frac{1}{10})s^2\mathbb{E}[\mathrm{Re}(\mathbf{h}^*\mathbf{a}_i\mathbf{a}_i^*\mathbf{x})|\mathbf{a}_i^*\mathbf{h}|^6]\\
	&&+9s^2\omega_i\mathbb{E}[\mathrm{Re}(\mathbf{h}^*\mathbf{a}_i\mathbf{a}_i^*\mathbf{x})^2|\mathbf{a}_i^*\mathbf{x}|^4]\\
	&\leq&(\omega_i+\frac{1}{2})^2\sqrt{\mathbb{E}[|\mathbf{a}_i^*\mathbf{h}|^8]\mathbb{E}[|\mathbf{a}_i^*\mathbf{x}|^8]}\\
	&&+(\omega_i-\frac{1}{10})^2s^4\mathbb{E}[|\mathbf{a}_i^*\mathbf{h}|^8]\\
	&&+9s^2\omega_i^2\sqrt{\mathbb{E}[|\mathbf{a}_i^*\mathbf{h}|^{12}]\mathbb{E}[|\mathbf{a}_i^*\mathbf{x}|^4]}\\
	&&+2(\omega_i+\frac{1}{2})(\omega_i-\frac{1}{10})\sqrt{\mathbb{E}[|\mathbf{a}_i^*\mathbf{h}|^{12}]\mathbb{E}[|\mathbf{a}_i^*\mathbf{x}|^4]}\\
	&&+6s\omega_i(2\omega_i+\frac{1}{2})\sqrt{\mathbb{E}[|\mathbf{a}_i^*\mathbf{h}|^{10}]\mathbb{E}[|\mathbf{a}_i^*\mathbf{x}|^6]}\\
	&&+6\omega_i(\omega_i-\frac{1}{10})s^3\sqrt{\mathbb{E}[|\mathbf{a}_i^*\mathbf{h}|^4]\mathbb{E}[|\mathbf{a}_i^*\mathbf{x}|^2]}\\
	&<&24(2\omega_i+\frac{1}{2})^2+24(\omega_i-\frac{1}{10})^2s^4\\
	&&+\sqrt{120\times42}\times6\omega_i(\omega_i-\frac{1}{10})s^3\\
	&&+\sqrt{16\times90}(9\omega^2+2(2\omega_i+\frac{1}{2})(\omega_i-\frac{1}{10})s^2\\
	&&+6\omega_i(2\omega_i+\frac{1}{2})\sqrt{5!\times3!}\\
	&<&500
\end{eqnarray*}
Thus let $\delta^2=m\mathrm{max}(9^2,500)=500m$ and $y=m/4$:
\begin{eqnarray*}
	\mathbb{P}(m\mu-\sum_{i=1}^{m}Y_i\geq\frac{m}{4})\leq e^{-2\gamma m}
\end{eqnarray*}
with $\gamma=1/2000$. With probability at least $1-e^{-2\gamma m}$, we have
\begin{eqnarray*}
	\frac{1}{m}Y_i(\mathbf{h},s)&\geq&(\omega_i+\frac{1}{2})\big(\frac{1}{2}+\frac{3}{2}\mathrm{Re}(\mathbf{x}^*\mathbf{h})^2\big)\\
	&+&6s\omega_i\mathrm{Re}(\mathbf{x}^*\mathbf{h})+3(\omega_i-\frac{1}{10})s^2-\frac{1}{4}
\end{eqnarray*}
when $\mathrm{Re}(\mathbf{x}^*\mathbf{h})>0$, from \cite{candes2015phase}, we can conclude that
\begin{eqnarray}
\frac{1}{m}Y_i(\mathbf{h},s)\geq \frac{3}{4}+\frac{3}{4}\mathrm{Re}(\mathbf{x}^*\mathbf{h})
\end{eqnarray}
when $\mathrm{Re}(\mathbf{x}^*\mathbf{h})<0$, we can also conclude the same inequality with $s\leq\sqrt{5/6}$.\\
Now that we prove it for a fixed vector, we should prove it for all $\mathbf{h}\in \mathbb{C}^n$ with $||\mathbf{h}||=1$.\\
Define $q(\mathbf{h})=\frac{1}{m}\sum_{i=1}^{m}p_i(\mathbf{h})-\frac{3}{4}\mathrm{Re}(\mathbf{x}^*\mathbf{h})^2$\\
where
\begin{eqnarray*}
	p_i(\mathbf{h})&=&(\omega_i+\frac{1}{2})\mathrm{Re}(\mathbf{h}^*\mathbf{a}_i\mathbf{a}_i^*\mathbf{x})^2\\
	&&+3\omega_is\mathrm{Re}(\mathbf{h}^*\mathbf{a}_i\mathbf{a}_i^*\mathbf{x})|\mathbf{a}_i^*\mathbf{h}|^2+(\omega_i-\frac{1}{10})s^2|\mathbf{a}_i^*\mathbf{h}|^4\\
\end{eqnarray*} 
Next, we will proof the Lipschitz property of $p_i(\mathbf{h})$, for any $\mathbf{u}$,$\mathbf{v}\in\mathbb{C}^n$,obeying $||\mathbf{u}||=||\mathbf{v}||=1$ we can have:
\begin{eqnarray}
\big|p_i(\mathbf{u})-p_i(\mathbf{v})\big|&=&(2\omega_i+\frac{1}{2})\big|\mathrm{Re}(\mathbf{u}^*\mathbf{a}_i\mathbf{a}_i^*\mathbf{x})^2\\
&&-\mathrm{Re}(\mathbf{v}^*\mathbf{a}_i\mathbf{a}_i^*\mathbf{x})^2\big|\nonumber\\
&&+3s\omega_i\Big|\mathrm{Re}(\mathbf{u}^*\mathbf{a}_i\mathbf{a}_i^*\mathbf{x})|\mathbf{a}_i^*\mathbf{u}|^2\nonumber\\
&&-\mathrm{Re}(\mathbf{v}^*\mathbf{a}_i\mathbf{a}_i^*\mathbf{x})|\mathbf{a}_i^*\mathbf{v}|^2\Big|\nonumber\\
&&+(\omega_i-\frac{1}{10})s^2\Big||\mathbf{a}_i^*\mathbf{u}|^4-|\mathbf{a}_i^*\mathbf{v}|^4\Big|\nonumber
\end{eqnarray} 
and
\begin{eqnarray}
&\big|&\mathrm{Re}(\mathbf{u}^*\mathbf{a}_i\mathbf{a}_i^*\mathbf{x})^2-\mathrm{Re}(\mathbf{v}^*\mathbf{a}_i\mathbf{a}_i^*\mathbf{x})^2\big|\nonumber\\
&=&\big|\mathrm{Re}\big((\mathbf{u}^*+\mathbf{v}^*)\mathbf{a}_i\mathbf{a}_i^*\mathbf{x}\big)\mathrm{Re}\big((\mathbf{u}^*-\mathbf{v}^*)\mathbf{a}_i\mathbf{a}_i^*\mathbf{x}\big)\big|\nonumber\\
&\leq&144n^2||\mathbf{u}-\mathbf{v}||
\end{eqnarray}
Similarly, we can have the inequalities below:
\begin{eqnarray}
\Big|\mathrm{Re}(\mathbf{u}^*\mathbf{a}_i\mathbf{a}_i^*\mathbf{x})|\mathbf{a}_i^*\mathbf{u}|^2-\mathrm{Re}(\mathbf{v}^*\mathbf{a}_i\mathbf{a}_i^*\mathbf{x})|\mathbf{a}_i^*\mathbf{v}|^2\Big|\nonumber\\
\leq108n^2||\mathbf{u}-\mathbf{v}||
\end{eqnarray}
\begin{eqnarray}
\Big||\mathbf{a}_i^*\mathbf{u}|^4-|\mathbf{a}_i^*\mathbf{v}|^4\Big|\leq72n^2||\mathbf{u}-\mathbf{v}||
\end{eqnarray}
Combining (6.12),(6.13),(6.14) into (6.11) and the bound of $\omega_i$, we can have:
\begin{eqnarray*}
	\big|p_i(\mathbf{u})-p_i(\mathbf{v})\big|&\leq&432n^2||\mathbf{u}-\mathbf{v}||+45n^2||\mathbf{u}-\mathbf{v}||\\
	&+&2n^2||\mathbf{u}-\mathbf{v}||\\
	&=&509n^2||\mathbf{u}-\mathbf{v}||
\end{eqnarray*}
As a result:
\begin{eqnarray*}
	\big|q(\mathbf{u})-q(\mathbf{v})\big|&\leq&\frac{1}{m}\sum_{i=1}^{m}|p_i(\mathbf{u})-p_i(\mathbf{v})|\\
	&&\frac{3}{4}|\mathrm{Re}(\mathbf{x}^*\mathbf{u})^2-\mathrm{Re}(\mathbf{x}^*\mathbf{v})^2|\\
	&\leq&509n^2||\mathbf{u}-\mathbf{v}||+\frac{3}{2}||\mathbf{u}-\mathbf{v}||\\
	&=&(509n^2+\frac{3}{2})||\mathbf{u}-\mathbf{v}||
\end{eqnarray*}
Therefore, for any $\mathbf{u},\mathbf{v}\in\mathbb{C}^n$,$||\mathbf{u}||=||\mathbf{v}||=1$ and $||\mathbf{u}-\mathbf{v}||\leq\eta:=\frac{1}{9000n^2}$, we can have:
\begin{eqnarray}
q(\mathbf{v})\geq q(\mathbf{u})-\frac{1}{16}
\end{eqnarray}
Let $\mathcal{N}_{\eta}$ be an $\eta-$ net for the unit sphere of $\mathbb{C}^n$ with cardinality satisfying $|\mathcal{N}_{\eta}|\leq(1+\frac{2}{\eta})^{2n}$. Applying with (5.10) with the union bound we can have for all $\mathbf{u}\in\mathcal{N}_{\eta}$
\begin{eqnarray}
\mathbb{P}\big(q(\mathbf{u})\geq\frac{3}{4}\big)&\geq& 1-|\mathcal{N}_{\eta}|e^{-2\gamma m}\nonumber\\
&\geq&1-(1+18000n^2)^ne^{-2\gamma m}\nonumber\\
&\geq&1-e^{-\gamma m}
\end{eqnarray}
Which holds by choosing $m$ such that $m\geq cnlogn$, where $c$ is a large constant. For any $\mathbf{h}$ on the unit sphere $\mathbb{C}^n$, there is a $\mathbf{u}\in\mathcal{N}_{\eta}$ such that $||\mathbf{h-\mathbf{u}}||\leq\eta$. Combining (6.15) and (6.16) we can have:
\begin{eqnarray*}
	q(\mathbf{h})&\geq&\frac{3}{4}-\frac{1}{16}>\frac{5}{8}\\
	&>&\frac{1}{m}Y_i(\mathbf{h},s)\geq(\frac{1}{8}+\frac{1}{2})+\frac{3}{4}\mathrm{Re}(\mathbf{x}^*\mathbf{h})^2
\end{eqnarray*}
which hold with probability at least $1-e^{-\gamma m}$ with $\alpha\geq8$
\subsection{Proof of the Local Smoothness Condition}
Through the knowledge of the operator, for any $\mathbf{z}\in\mathbb{E}$ To prove (6.1) is equivalent to prove that for all $\mathbf{u}\in \mathbb{C}^n$ obeying $||\mathbf{u}||=1$, we will have 
\begin{eqnarray*}
	|(\nabla f(\mathbf{z}))^*\mathbf{u}|^2
	&\leq&\beta(\frac{(1-\delta)}{4}dist^2(\mathbf{z},\mathbf{x})\\
	&&+\frac{1}{10m}\sum_{i=1}^{m}|\mathbf{a}_i^*(\mathbf{z}-e^{i\phi(\mathbf{z})}\mathbf{x})|^4)
\end{eqnarray*}
So we define:
\begin{eqnarray*}
	g(\mathbf{h},\mathbf{\omega},s)&=&\frac{1}{m}\sum_{i=1}^{m}2\omega_i\mathrm{Re}(\mathbf{h}^*\mathbf{a}_i\mathbf{a}_i^*\mathbf{x})\mathrm{Re}(\mathbf{\nu}^*\mathbf{a}_i\mathbf{a}_i^*\mathbf{x})\\
	&&+\frac{1}{m}\sum_{i=1}^{m}s\omega_i|\mathbf{a}_i^*\mathbf{h}|^2\mathrm{Re}(\mathbf{\nu}^*\mathbf{a}_i\mathbf{a}_i^*\mathbf{x})\\
	&&+\frac{1}{m}\sum_{i=1}^{m}2s\omega_i\mathrm{Re}(\mathbf{h}^*\mathbf{a}_i\mathbf{a}_i^*\mathbf{x})\mathrm{Re}(\mathbf{\nu}^*\mathbf{a}_i\mathbf{a}_i^*\mathbf{h})\\
	&&+\frac{1}{m}\sum_{i=1}^{m}s^2\omega_i|\mathbf{a}_i^*\mathbf{h}|^2\mathrm{Re}(\mathbf{\nu}^*\mathbf{a}_i\mathbf{a}_i^*\mathbf{x})\\
\end{eqnarray*}
where $\mathbf{h}=e^{-j\phi(\mathbf{z})}\mathbf{z}-\mathbf{x}$ and $\nu=e^{-j\phi(\mathbf{z})}\mathbf{u}$, Thus to prove Local Smoothness Condition, it is sufficient to prove that:
\begin{eqnarray*}
	|g(\mathbf{h},\nu,1)|^2&\leq&\beta(\frac{1+\delta}{4}||\mathbf{h}||^2\\
	&&+\frac{1}{10m}\sum_{i=1}^{m}|\mathbf{a}_i^*(\mathbf{z}-e^{i\phi(\mathbf{z})}\mathbf{x})|^4)
\end{eqnarray*}
holds for all $\mathbf{h}$ and $\nu$ satisfying $\mathrm{Im}(\mathbf{h}^*\mathbf{x})=0$,$||\mathbf{h}||\leq\epsilon$ and $||\omega||=1$. Using the same idea in proving Local Curvature Condition, we merely need to prove for $\mathbf{h}$ and $\mathbf{\nu}$ satisfying $\mathrm{Im}(\mathbf{h}^*\mathbf{x})=0$, $||\mathbf{h}||=||\mathbf{\nu}||=1$ and $\forall s:0\leq s\leq\epsilon$
\begin{eqnarray*}
	|g(\mathbf{h},\nu,s)|^2\leq\beta(\frac{1+\delta}{4}+\frac{s^2}{10m}\sum_{i=1}^{m}|\mathbf{a}_i^*(\mathbf{z}-e^{i\phi(\mathbf{z})}\mathbf{x})|^4)
\end{eqnarray*} 
By using the inequality $(a+b+c)^2\leq3(a^2+b^2+c^2)$
\begin{eqnarray*}
	|g(\mathbf{h},\mathbf{\nu},s)|^2&\leq&\Big|\frac{1}{m}\sum_{i=1}^{m}2\omega_i|\mathbf{h}^*\mathbf{a}_i||\mathbf{\nu}^*\mathbf{a}_i||\mathbf{a}_i^*\mathbf{x}|^2\\
	&&+\frac{1}{m}\sum_{i=1}^{m}3s\omega_i|\mathbf{a}_i^*\mathbf{h}|^2|\mathbf{\nu}^*\mathbf{a}_i||\mathbf{a}_i^*\mathbf{x}|\\
	&&+\frac{1}{m}\sum_{i=1}^{m}s^2\omega_i|\mathbf{a}_i^*\mathbf{h}|^3|\mathbf{\nu}^*\mathbf{a}_i|\Big|^2\\
	&\leq&\frac{10}{3}\Big|\frac{1}{m}\sum_{i=1}^{m}2|\mathbf{h}^*\mathbf{a}_i||\mathbf{\nu}^*\mathbf{a}_i||\mathbf{a}_i^*\mathbf{x}|^2\Big|^2\\
	&&+\frac{10}{3}\Big|\frac{1}{m}\sum_{i=1}^{m}3s|\mathbf{a}_i^*\mathbf{h}|^2|\mathbf{\nu}^*\mathbf{a}_i||\mathbf{a}_i^*\mathbf{x}|\Big|^2\\
	&&+\frac{10}{3}\Big|\frac{1}{m}\sum_{i=1}^{m}s^2|\mathbf{a}_i^*\mathbf{h}|^3|\mathbf{\nu}^*\mathbf{a}_i|\Big|^2\\
	&=&\frac{10}{3}(4I_1+9s^2I_2+\frac{m^2}{s^4}I_3)
\end{eqnarray*}
Through Cauchy-Schwarz inequality and Corollary 7.6 in\cite{candes2015phase} we can have:
\begin{eqnarray*}
	I_1&\leq&(\frac{1}{m}\sum_{i=1}^{m}(|\mathbf{x}^*\mathbf{a}_i||\mathbf{\nu}^*\mathbf{a}_i|)^2)(\frac{1}{m}\sum_{i=1}^{m}(|\mathbf{x}^*\mathbf{a}_i||\mathbf{h}^*\mathbf{a}_i|)^2)\\
	&\leq&(2+\delta)^2
\end{eqnarray*}
Similarly, we have
\begin{eqnarray*}
	I_2&\leq&(\frac{1}{m}\sum_{i=1}^{m}|\mathbf{a}_i^*\mathbf{h}|^4)(\frac{1}{m}\sum_{i=1}^{m}|\mathbf{a}_i^*\mathbf{\nu}|^2|\mathbf{a}_i^*\mathbf{x}|^2)\\
	&\leq&\frac{2+\delta}{m}\sum_{i=1}^{m}|\mathbf{a}_i^*\mathbf{h}|^4
\end{eqnarray*}
For $I_3$, using Cauchy-Schwarz inequality and with Lemma 7.8 in\cite{candes2015phase} we can have:
\begin{eqnarray*}
	I_3&\leq&(\frac{1}{m}\sum_{i=1}^{m}|\mathbf{a}_i^*\mathbf{h}|^3\mathop {\mathrm{max}}\limits_{i}||\mathbf{a}_i||)^2\\
	&\leq&6n(\frac{1}{m}\sum_{i=1}^{m}|\mathbf{a}_i^*\mathbf{h}|^3)^2\\
	&\leq&6n(\frac{1}{m}\sum_{i=1}^{m}|\mathbf{a}_i^*\mathbf{h}|^4)(\frac{1}{m}\sum_{i=1}^{m}|\mathbf{a}_i^*\mathbf{h}|^2)\\
	&\leq&6n(1+\delta)\frac{1}{m}|\mathbf{a}_i^*\mathbf{h}|^4
\end{eqnarray*}
As a result:
\begin{eqnarray*}
	|g(\mathbf{h},\mathbf{\nu},s)|^2&\leq&\frac{40}{3}(2+\delta)^2+30(2+\delta)s^2\frac{1}{m}\sum_{i=1}^{m}|\mathbf{a}_i^*\mathbf{h}|^4\\
	&~&+20s^4n(1+\delta)\frac{1}{m}|\mathbf{a}_i^*\mathbf{h}|^4\\
	&\leq&\beta(\frac{1+\delta}{4}+\frac{s^2}{10m}\sum_{i=1}^{m}|\mathbf{a}_i^*\mathbf{h}|^4)
\end{eqnarray*}
which holds:\\
$\beta\geq max\Big(\frac{160}{3}\frac{(2+\delta)^2}{(1+\delta
	)},300(2+\delta)+200\epsilon^2n(1+\delta)\Big)$\\
In the theorem, the $\delta\leq0.01$ and $\epsilon$ can be choose to be $\frac{1}{8}$ and $\frac{1}{8\sqrt{n}}$. Then we conclude our proof.
\end{document}